\documentclass[epj]{webofc}
\usepackage[varg]{txfonts}   % Web of Conferences font
\wocname{EPJ Web of Conferences}
\woctitle{ICNFP 2017}
\begin{document}
\selectlanguage{english}
\title{Probing the tricritical endpoint of QCD phase diagram at NICA-FAIR energies}
%
% subtitle is optional
%
%%%\subtitle{Do you have a subtitle?\\ If so, write it here}

\author{K. A. Bugaev\inst{1} %\fnsep\thanks{\email{Bugaev@th.physik.uni-frankfurt.de}} 
\and
        A. I. Ivanytskyi\inst{1}
        \and
         V. V. Sagun\inst{1,2}, G. M. Zinovjev\inst{1} 
        \and
        E. G. Nikonov\inst{3}
        \and
        R. Emaus\inst{4}
        \and
        L. V.  Bravina\inst{4}
        \and
        E. E.  Zabrodin\inst{4, 5, 6}
        \and
        A. V.  Taranenko\inst{6}
}

\institute{Bogolyubov Institute for Theoretical Physics of the National Academy of Sciences of Ukraine, 03680 Kiev,  Ukraine
\and
Centro de Astrof\'{\i}sica e Gravita\c c\~ao  - CENTRA,
Departamento de F\'{\i}sica, Instituto Superior T\'ecnico,
Universidade de Lisboa, 1049-001 Lisboa, Portugal
\and
 Laboratory for Information Technologies, Joint Institute for Nuclear Research,   Dubna 141980, Russia
 \and
Department of Physics, University of Oslo, PB 1048 Blindern, N-0316 Oslo, Norway 
\and
Skobeltzyn Institute of Nuclear Physics, Moscow State University, 119899 Moscow, Russia
\and
National Research Nuclear University ``MEPhI'' (Moscow Engineering Physics Institute), 115409 Moscow, Russia
}

\abstract{
In this contributions we discuss the novel version of  hadron resonance 
gas model which is based on the induced surface tension concept. 
Also we present  new arguments in favor of a hypothesis that the chiral 
symmetry restoration transition in central nuclear collisions may occur 
at the center of mass energies 4.3-4.9 GeV and that the deconfinement 
phase transition may occur at the center of mass energies 8.8-9.2 GeV.
These arguments are based on the unique thermostatic properties of the 
mixed phase and the ones of an exponential mass spectrum of hadrons.
}
\maketitle
\section{Introduction}
\label{intro_Bugaev}

During more than thirty years of  searches for  the deconfinement and 
for the chiral symmetry restoration (CSR) phase transitions (PT) in 
heavy ion collisions (HIC) there were made many interesting discoveries 
\cite{QM17}, but until recently the situation with the reliable signals 
was rather controversial. Although such peculiar  irregularities in 
excitation functions of experimental data,  known in the literature as 
the Kink \cite{Kink}, the Strangeness Horn \cite{Horn} and the Step 
\cite{Step}, motivated the experimental program at CERN SPS to lower 
the collision energy in order  to locate the onset of deconfinement, 
an absence of firm  theoretical explanation of these irregularities was 
always the source of natural doubts about their relation to 
deconfinement PT. 
After a discovery of the step-like structure of  the inverse slope 
parameter of transverse momentum spectra of K$^+$ and K$^-$ mesons in 
proton-proton collisions \cite{StepPP} at the same collision energies 
as in HIC, the whole logic of claims made in  Ref.~\cite{Gazd_rev:10} 
was destroyed. Indeed, the major argument of Ref.~\cite{Gazd_rev:10} 
that  the Kink, Strangeness Horn and Step observed in HIC are the 
signals of deconfinement is based on the fact that the corresponding  
quantities observed in proton-proton collisions behave entirely 
different. Therefore, after the similarity of transverse momentum 
spectra of K$^+$ and K$^-$ mesons measured in HIC and in proton-proton 
collision was found, it became evident that without firm theoretical 
back-up all speculations of Ref.~\cite{Gazd_rev:10} about the Kink, 
Strangeness Horn and Step as the signals of the onset of deconfinement 
are not trustworthy. 

It is necessary to stress that such a situation is not surprising, 
since  in HIC we are dealing with finite systems for which there is no 
rigorous theory of first order PT of the liquid-gas type. Although some 
progress in formulation of the finite volume analog of phases for 
liquid-gas PT in finite static systems was achieved  on the basis of 
exactly solvable models \cite{SMM05,SMM07}, presently it is unclear how 
to apply these results  to the fast expanding systems created in HIC.  
It is, therefore, clear that any new signal of the first order 
liquid-gas PT which can be applied to finite systems is very important 
for a completion of the HIC programs which are performed presently at 
SPS CERN, LHC  CERN and RHIC BNL  and which are planned  for the 
nearest future at NICA JINR  and FAIR GSI. Therefore, here we would 
like to present a few irregularities found during last three years 
\cite{Bugaev:2015,Bugaev:2016a,Bugaev:2016b,Bugaev:2017a,Bugaev:2017b} 
in symmetric HIC at chemical freeze-out (CFO) and to briefly discuss 
their relation to deconfinement PT and to CSR PT. 

The work is organized as follows. In Sect.~2 we remind the basic 
elements of the newest hadron resonance gas model (HRGM) 
\cite{IST1,IST2,IST3} based on the concept of induced surface tension 
\cite{IST14}. The new signals of CSR and deconfinement PT are discussed 
in Sect.~3, whereas our conclusions are summarized in Sect.~4.

\section{ HRGM with multicomponent hard-core repulsion}

Traditionally, the HRGM  
\cite{Andronic:05,Oliinychenko:12,HRGM:13,SFO:13,Sagun,Sagun2,BugaevUJP16} 
is used to determine the parameters of CFO from the measured hadronic 
yields. Presently its version with the multicomponent hard-core 
repulsion between hadrons 
 \cite{Oliinychenko:12,HRGM:13,SFO:13,Sagun,Sagun2,IST1} provides the  
best description of all independent hadronic multiplicity rations
measured in the central HIC at the center of mass energies 
$\sqrt s_{NN}=2.7, 3.1, 3.8, 4.3, 4.9, 6.3, 7.6,  8.8,  9.2, 12.3, 
17.3, 62.4, 130, 200, 2760$~GeV. There are three main reasons to employ 
the HRGM as the equation of state (EoS) of hadronic matter. First, 
it is well known that for temperatures below 170 MeV and low baryonic 
charge densities the  mixture of stable hadrons and their resonances
whose interaction is taken into account by the quantum  second virial 
coefficients behaves as the mixture of nearly ideal gases of stable 
particles which, however, includes both the hadrons and their 
resonances, but taken with their averaged masses \cite{Raju}. As it was 
shown in Ref.~\cite{Raju}, the main reason for  such a behavior is 
related to  an almost complete cancellation between the attraction  and 
repulsion contributions in the quantum second virial coefficients. 
Therefore, the remaining deviation from the ideal gas (a weak repulsion) 
is usually described by the classical second virial coefficients. The 
second reason to use the HRGM as the hadronic matter EoS is that in 
this case its pressure will not exceed the one of quark-gluon plasma 
which, nevertheless, may happen, if the hadrons are treated as the 
mixture of ideal gases \cite{IST1,Satarov10}. Finally, the third reason 
to employ the HRGM is the practical one: due to the fact that hard-core 
repulsion is the contact interaction, the energy per particle of such 
an EoS coincides with the one of  an ideal gas, even for quantum 
particles \cite{IST3}. As a consequence, during the subsequent 
evolution of the system after the CFO to the kinetic freeze-out 
\cite{KABkinFO1,KABkinFO2} one should not somehow ``convert'' the 
potential energy of interacting particles into their kinetic energy 
and/or into the masses of newly born particles. These reasons allow one 
not only to consider the HRGM as an extension of the famous statistical 
bootstrap model \cite{Hagedorn} supplemented by the hard-core repulsion 
which, in addition, has a truncated hadronic mass spectrum, but to 
effectively use it to describe the  hadronic multiplicities measured in 
the HIC experiments. 

Despite many valuable results obtained with the HRGM during last years, 
the hard-core radii are well established at the moment for the most 
abundant hadrons only, namely for pions ($R_\pi \simeq 0.15$ fm), for 
lightest  K$^\pm$-mesons ($R_K \simeq 0.395$ fm), for nucleons 
($R_p \simeq 0.365$ fm) and for lightest (anti)$\Lambda$-hyperons 
($R_\Lambda \simeq 0.085$ fm) \cite{IST1,IST2}. 
Nevertheless, we hope that the new data of high quality which are 
expected to be measured during the Beam Energy Scan II at RHIC BNL 
(Brookhaven) \cite{RHIC17}, and at the accelerators of new generation, 
i.e. at NICA JINR (Dubna) \cite{NICA} and FAIR GSI (Darmstadt) 
\cite{FAIR}, will help us to determine the hard-core radii of other 
measured hadrons with high accuracy. However, the traditional 
multicomponent HRGM is not suited for such a purpose, since for $N$ 
different hard-core radii it is necessary to find a solution of $N$ 
transcendental equations. Hence, a further increase of the number of 
hard-core radii (i.e., $N\sim 100$, corresponding to the various 
hadronic species created in a collision) will lead to a huge increase 
of computational time and in this way it will destroy the main 
attractive feature of the HRGM, i.e., its simplicity. To overcome this 
problem the novel HRGM based one the concept of the induced surface 
tension (IST) \cite{IST14} was recently developed in 
Refs.~\cite{IST1,IST2,IST3}.

It is a system of coupled  equations for the pressure $p$ and the 
induced surface tension coefficient $\Sigma$
\begin{eqnarray}
\label{EqI}
p &=& \sum_{k=1}^N  p_k =  T \sum_{k=1}^N \phi_k \exp \left[ 
\frac{\mu_k}{T} - \frac{4}{3}\pi R_k^3 \frac{p}{T} - 4\pi R_k^2 
\frac{\Sigma}{T} \right]
\,, \\
\label{EqII}
\Sigma &=& \sum_{k=1}^N  \Sigma_k =  T \sum_{k=1}^N R_k \phi_k \exp 
\left[ \frac{\mu_k}{T} - \frac{4}{3}\pi R_k^3 \frac{p}{T} - 4\pi R_k^2 
\alpha \frac{\Sigma}{T} \right] \,,\\
\label{EqIII}
\mu_k &=& \mu_B B_k + \mu_{I3} I_{3k} + \mu_S S_k \,,
\end{eqnarray}
where $\alpha = 1.245$, and $\mu_B$, $\mu_S$, $\mu_{I3}$ are the 
baryonic, the strangeness, and the third projection of the isospin 
chemical potential, respectively. Here $B_k$, $S_k$, $I_{3k}$, $m_k$ and 
$R_k$ denote, respectively, the corresponding charges, mass, and 
hard-core radius of the $k$-th hadronic species. The sums in 
Eqs.~(\ref{EqI}) and (\ref{EqII}) run over all hadronic species; their 
corresponding antiparticles are considered as independent species and, 
hence, $p_k$ and $\Sigma_k$ are, respectively, the partial pressure and 
the partial induced surface tension coefficient of the $k$-th hadronic 
species. 

The one-particle thermal density $\phi_k$ in Eqs.~(\ref{EqI}) and 
(\ref{EqII}) accounts for the Breit-Wigner mass attenuation and is 
written in the Boltzmann approximation (the quantum one is given in 
\cite{IST3})
\begin{eqnarray}
\label{EqIV}
\phi_k = g_k  \gamma_S^{|s_k|} \int\limits_{M_k^{Th}}^\infty  \,  
\frac{ d m}{N_k (M_k^{Th})} 
\frac{\Gamma_k}{(m-m_{k})^{2}+\Gamma^{2}_{k}/4} 
\int \frac{d^3 p}{ (2 \pi)^3 }   \exp \left[ -
\frac{ \sqrt{p^2 + m^2} }{T} \right] \,,
\end{eqnarray}
where $g_k$ is the degeneracy factor of the $k$-th hadronic species,
$\gamma_S$ is the strangeness suppression factor \cite{Rafelski}, 
$|s_k|$ is the number of valence strange quarks and antiquarks in this 
hadron species, $\displaystyle {N_k (M_k^{Th})} \equiv \int
\limits_{M_k^{Th}}^\infty \frac{d m \, \Gamma_k}{(m-m_{k})^{2}+
\Gamma^{2}_{k}/4} $ denotes 
a normalization factor, while $M_k^{Th}$ corresponds to the decay 
threshold mass of the $k$-th hadronic species and $\Gamma_k$ denotes 
its width.

To apply the system of Eqs.~(\ref{EqI}), (\ref{EqII}), and (\ref{EqIII}) 
to study nuclear collisions it should be supplemented by 
the strange charge conservation law, which is equivalent to a vanishing 
net strangeness density
\begin{eqnarray}
\label{EqV}
\rho_S\equiv\frac{\partial p}{\partial \mu_S}= \sum_k S_k\, \rho_k=0 \,,
\end{eqnarray}
where $\rho_k$ is density of hadrons of sort $k$ given by the system
of equations
\begin{eqnarray}\label{EqVIA}
&&\rho_k \,\,~\equiv~\, \, \frac{\partial  p}{\partial \mu_k} = \frac{1}{T} \cdot \frac{p_k \, a_{22} 
- \Sigma_k \, a_{12}}{a_{11}\, a_{22} - a_{12}\, a_{21} } \,,\\
&&a_{11} ~=~ 1 + \frac{4}{3}\, \pi \sum_k  R_k^3 \frac{p_k}{T} \,, \, \quad  a_{12} ~=~ 4 \pi \sum_k R_k^2 \frac{p_k}{T} \, ,\\
&&a_{22} ~=~ 1 + 4 \pi \alpha \sum_k  R_k^2  \frac{\Sigma_k}{T} \,, \quad a_{21} ~=~ \frac{4}{3} \pi \sum_k R_k^3\frac{\Sigma_k}{T} \,.
\end{eqnarray}

Note that in contrast to the usual multicomponent HRGM formulations to 
determine the particle number densities $\{ \rho_k \}$ one needs to 
solve only a system of  three equations, namely Eqs. (1), (2) and (5), 
irrespective to the number of different hard-core radii in the model. 
Therefore, we believe that IST EoS defined by system (1)-(5) is 
perfectly suited for the analysis of all hadronic multiplicities 
which will be measured in the nearest future at RHIC, NICA and FAIR. 

Another great advantage of the IST EoS is its validity up to the 
packing fractions $\eta \equiv \sum_k \frac{4}{3}\pi R_k^3 \rho_k 
\simeq  0.2-0.22$ \cite{IST1,IST2,IST3}, i.e., at the particle number 
densities for which the traditional HRGM  based on the Van der Waals 
approximation 
\cite{Andronic:05,Oliinychenko:12,HRGM:13,SFO:13,Sagun,Sagun2} 
is absolutely incorrect.  

Using the particle number density (\ref{EqVIA}) of $k$-th sort of 
hadrons one can determine the thermal $N_k^{th} =V \rho_k$ ($V$ is the 
effective volume at CFO) and the total multiplicities $N_k^{tot}$.
The latter should account for the hadronic decays after the CFO and 
then the ratio of total hadronic multiplicities becomes 
\begin{equation}
\label{EqIX}
\frac{N^{tot}_k}{N^{tot}_j}=
\frac{\rho_k+\sum_{l\neq k}\rho_l\, Br_{l\rightarrow k}}{\rho_j+
\sum_{l\neq j}\rho_l \, Br_{l\rightarrow j}}\,,
\end{equation}
where $Br_{l\rightarrow k}$ is the branching ratio, i.e., a probability 
of particle $l$ to decay strongly into a particle $k$. More details on 
the fitting procedure of experimental data with the HRGM can be found in  
\cite{Sagun}.

%%%%%%%%%%%%%%% Section 2

\section{New Signals of QCD Phase Transitions}
\label{sect2_Bugaev}

\begin{figure}[t]
% Use the relevant command for your figure-insertion program
% to insert the figure file.
\centering
\mbox{\includegraphics[width=70mm,clip]{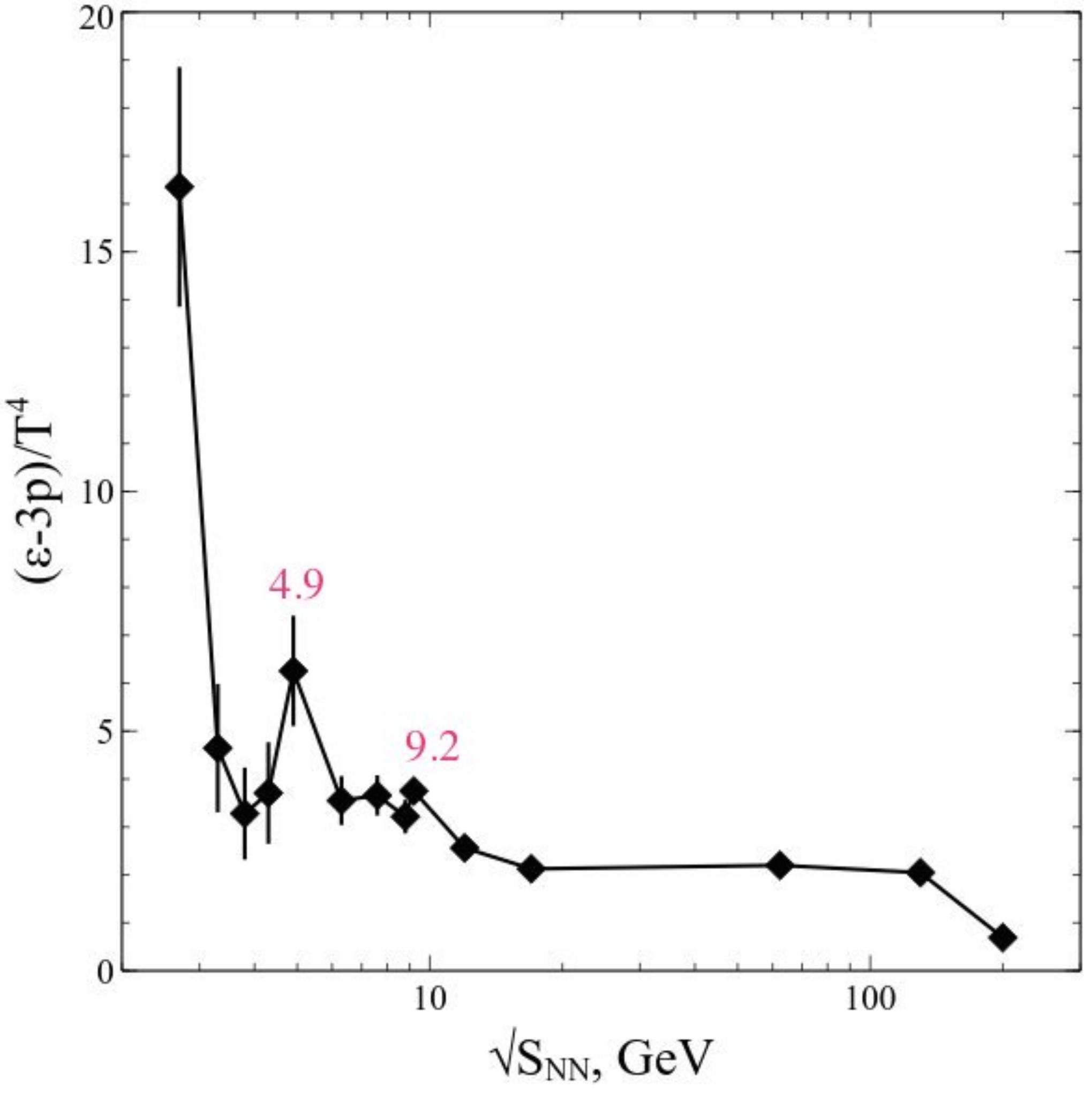}
\hspace*{0.2mm}
\includegraphics[width=70mm,clip]{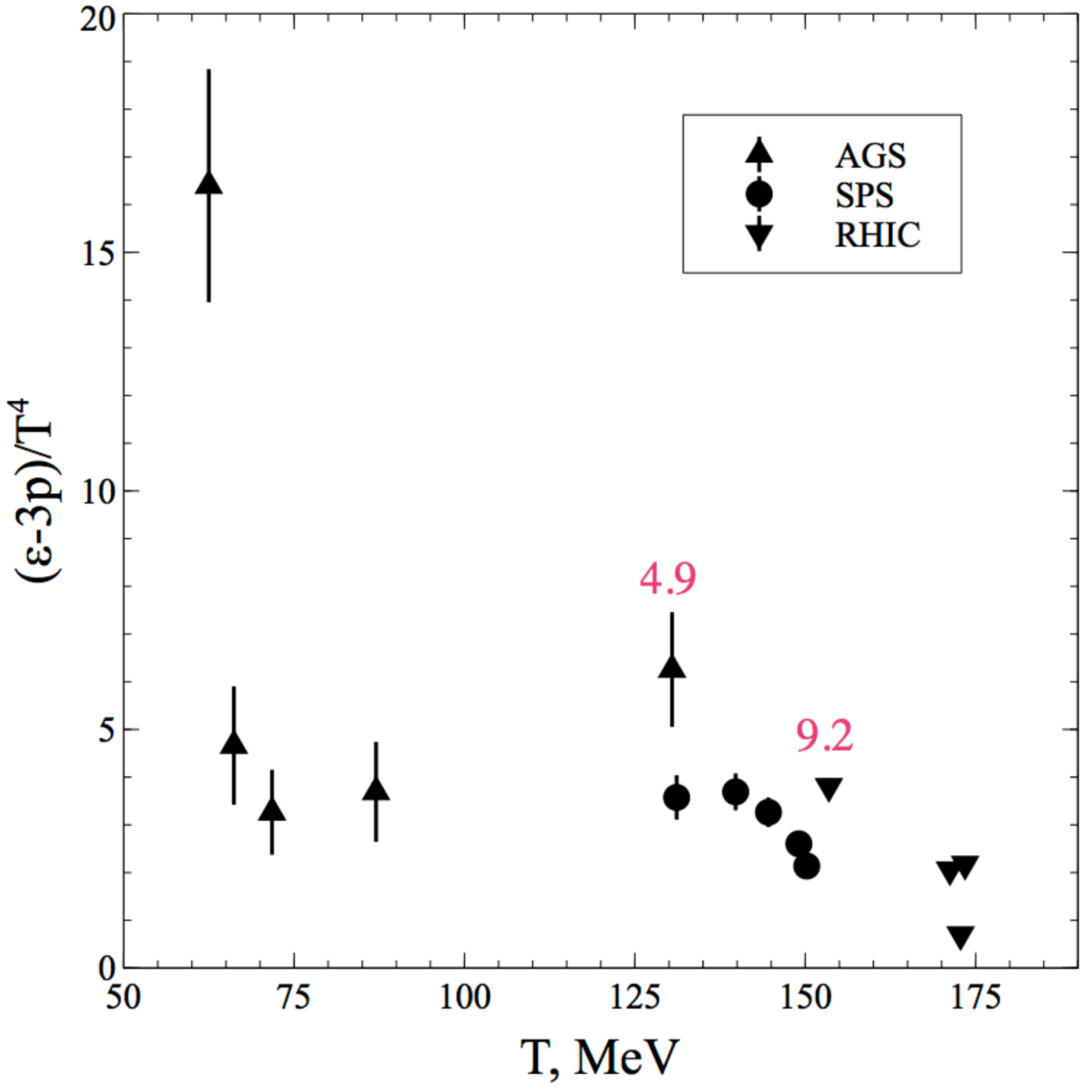}}
\caption{Trace anomaly dependence  on the center of mass collision 
energy (left) and on  the CFO temperature (right). The numbers above 
the symbols correspond to a particular value of center of mass collision 
energy. In the right panel the points $\sqrt{s_{NN}} =4.9$~GeV and 
$\sqrt{s_{NN}} =9.2$~GeV are well separated from the neighboring ones. 
}
\label{Bugaev_fig1}       % Give a unique label
\end{figure}

\begin{figure}[h]
% Use the relevant command for your figure-insertion program
% to insert the figure file.
\centering
\mbox{\includegraphics[width=90mm,clip]{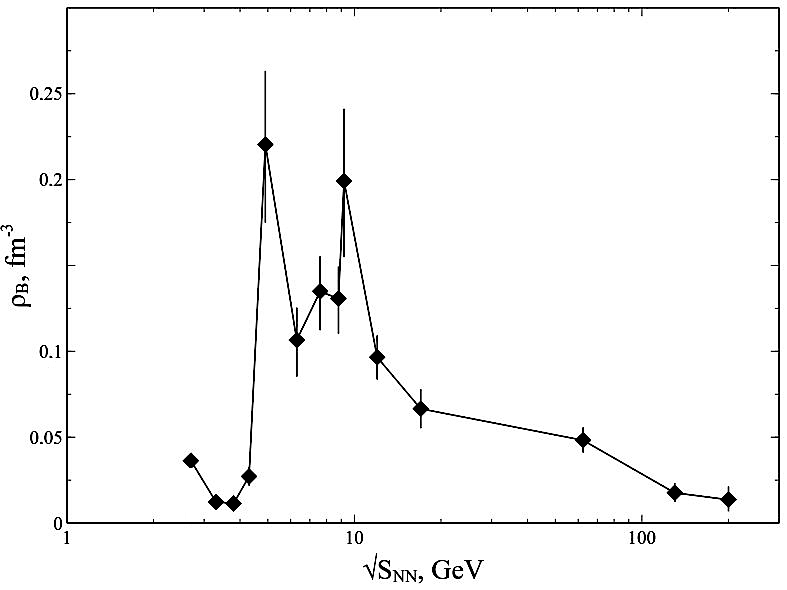}}
\caption{Collision energy dependence of the baryonic charge density 
found by the IST EoS at CFO. }
\label{Bugaev_fig2}       % Give a unique label
\end{figure}

Using the multicomponent HRGM it was possible to reveal a few 
irregularities of thermodynamic quantities  observed at CFO  and to 
relate them to two QCD  phase transitions. The most remarkable 
irregularities include two sets of correlated quasi-plateaus found in  
\cite{Bugaev:2015,Bugaev:2016a,Bugaev:2016b} which are located at the  
collision energy ranges $\sqrt{s_{NN}} \simeq 3.8-4.9$~GeV  and  
$\sqrt{s_{NN}} \simeq  7.6-9.2$~GeV, and two  peaks of trace anomaly 
$\delta = \frac{(\epsilon - 3p)}{T^4}$ (here $\epsilon$, $p$ and $T$ 
denote, respectively, the energy density of the system, its pressure 
and temperature) observed at the maximal energy of each set of 
quasi-plateaus \cite{Bugaev:2016b,Bugaev:2017a}. Figure~\ref{Bugaev_fig1}  
shows the trace anomaly $\delta$ as a function of center-of-mass 
collision energy and as a function of the CFO temperature
obtained by the IST EoS \cite{Bugaev:2017a}. As one can see from 
Fig.~\ref{Bugaev_fig2} the baryonic charge  density at CFO exhibits two 
sharp peaks which are located exactly at the collision energies of the 
trace anomaly peaks \cite{Bugaev:2017a}, i.e. at $\sqrt{s_{NN}}=4.9$~GeV 
and $\sqrt{s_{NN}} = 9.2$~GeV. 

Although the set of low energy quasi-plateaus was predicted a long time 
ago \cite{KAB:89a,KAB:90} as a manifestation of  the  anomalous 
thermodynamic properties of  quark-gluon-hadron mixed phase, the 
interpretation of the high energy set of quasi-plateaus was difficult, 
since the generalized shock adiabat model \cite{KAB:89a,KAB:90}  
which  predicted such  quasi-plateaus can be applied to the collision 
energies $\sqrt{s_{NN}} >  8$~GeV qualitatively only.  
Thus, the first hints in favor of two QCD phase transitions were found 
in 2014 in \cite{Bugaev:2015}, but the self-consistent interpretation 
of these irregularities was worked out only in 2017 in 
\cite{Bugaev:2017a}. This success was achieved  after finding out in 
Ref.~\cite{Bugaev:2017a} the number of effective degrees of freedom of 
the phase formed at the collision energies $\sqrt{s_{NN}} = 4.9-9.2$~GeV 
and after explaining in Ref.~\cite{Bugaev:2017a} the peculiar collision 
energy dependence of the strangeness enhancement factor $\gamma_s$. 
Since the number of effective degrees of freedom was already discussed 
in Ref. \cite{Bugaev:2017b}, here we discuss the novel signals of two 
QCD phase transitions related to the $\gamma_s$ factor. 

\begin{figure}[t]
% Use the relevant command for your figure-insertion program
% to insert the figure file.
\centering
\mbox{\includegraphics[width=70mm,height=60mm,clip]{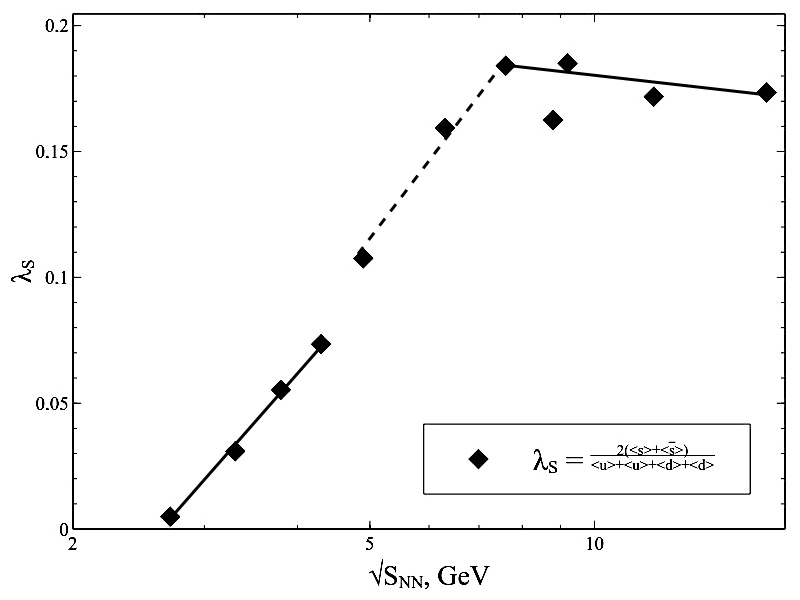}\hspace*{0.2mm}\includegraphics[width=70mm,height=61mm,clip]{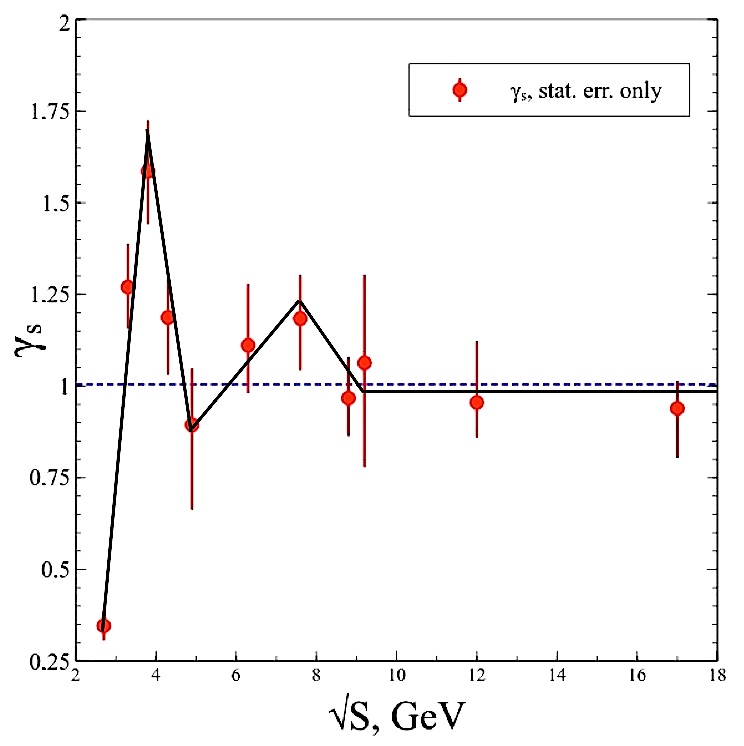}}
\caption{Comparison of the collision energy dependence of the modified 
Wroblewski factor $\lambda_s$ (left panel) and 
the strangeness suppression factor $\gamma_s$ (right panel).}
\label{Bugaev_fig3}       % Give a unique label
\end{figure}

For more than three decades the HIC community is discussing the relation 
of strange charge irregularities to a deconfinement PT, but until now 
there is no complete understanding of such a relation. To demonstrate a 
single aspect of this problem, let us compare the left and the right 
panels of  Fig.~\ref{Bugaev_fig3}.
From the left panel of Fig.~\ref{Bugaev_fig3} one can see that up to 
the collision energy $\sqrt s_{NN} = 8.8$~GeV there is a monotonic 
increase of the modified Wroblewski factor $\lambda_s$ \cite{Wrobl} 
which was defined in Ref.~\cite{Bugaev:2017a} as 
\begin{eqnarray}\label{EqX}
\lambda_s \equiv \frac{2 \sum\limits_{n} (N_n^S + N_n^{\bar{S}}) \rho_n }
{ \sum\limits_{n} (N_n^u +  N_n^{\bar{u}} +
N_n^d +  N_n^{\bar{d}}) \rho_n } \,,
\end{eqnarray}
where in the numerator $N_n^S$ and $N_n^{\bar{S}}$ denote, respectively, 
the number of strange valence quarks and antiquarks in the hadron of 
sort $n$, whereas  $N_n^u$ and $N_n^d$ in the denominator denote, 
respectively, the number of $u$ and $d$ valence quarks in it (for 
antiquarks we used $N_n^{\bar{u}}$ and $N_n^{\bar{d}}$). Note that the 
denominator in Eq.~(\ref{EqX}) differs from the traditional Wroblewski 
factor \cite{Wrobl} because it accounts for the  number of $u$ and $d$  
valence quarks and antiquarks, and not only the ones which are  paired 
to their valence  antiquarks (quarks). 
This quantity is more convenient and has a simple physical meaning: 
$\lambda_s$  (\ref{EqX}) can be treated as an asymmetry of the strange 
(anti)quark production compared to  the  $u$ and $d$ quarks. Also, the  
$\lambda_s$ factor is directly related to the concentration of 
(anti)strange quarks in the system, i.e  
\begin{eqnarray}\label{EqXI}
\nu_s = \frac{\sum\limits_{n} (N_n^S + N_n^{\bar{S}})  \rho_n }{\sum\limits_{n} (N_n^u +  N_n^{\bar{u}} + N_n^d +  N_n^{\bar{d}} + N_n^S + N_n^{\bar{S}}) \rho_n} \equiv \frac{\lambda_s }{2 + \lambda_s }\,.
\end{eqnarray}

On the other hand the right panel of Fig.~\ref{Bugaev_fig3} demonstrates 
a very complicated behavior of the $\gamma_s$ factor right for collision 
energies $\sqrt s_{NN} < 8.8$~GeV, while for $\sqrt s_{NN} \ge 8.8$~GeV 
the strangeness demonstrates a full chemical equilibrium, i.e., 
$\gamma_s=1$. Such a behavior was first found in 2013, when the 
$\gamma_s$ factor was used in the multicomponent HRGM of 
Ref.~\cite{SFO:13}, but until recently its reason for such a behavior 
remained a hard puzzle. 
First we discuss the collision energies above $8.8$~GeV. As it is argued 
in \cite{Bugaev:2017a} the fact that for these energies $\gamma_s=1$ 
within the error bars can be naturally explained by the formation of the 
quark-gluon bags with the exponential mass spectrum proposed by 
R.~Hagedorn \cite{Hagedorn}. As it was predicted in \cite{Thermostat1} 
and later on shown numerically in \cite{Hstate1,Hstate2}, the 
exponential mass spectrum acts as a perfect thermostat and a perfect 
particle reservoir, i.e., all particles which appear from such a bag at 
its hadronization will inevitably be born in a state of full thermal 
and chemical equilibrium with it \cite{Thermostat1,Hstate1,Hstate2}.  
Therefore, most probably, the collision energy 
$\sqrt s_{NN} \simeq 8.8-9.2$~GeV corresponds to the deconfinement PT to 
QGP \cite{Bugaev:2017a,Bugaev:2017b}. More arguments in favor of this 
conclusion are given in  \cite{Bugaev:2017a,Bugaev:2017b}.

As argued in \cite{Bugaev:2017a} the dale of $\gamma_s$ factor, i.e., 
$\gamma_s=1$ at $\sqrt s_{NN} = 4.9$~GeV, is directly related to the 
1-st order PT. Indeed, in Ref.~\cite{Thermostat1} a few examples of the 
explicit thermostats and explicit particle reservoirs are discussed and 
it is argued that under the constant pressure condition the mixed phase 
of the 1-st order PT, i.e., two pure phases being in a full thermal and 
chemical equilibrium with each other, represent both a thermostat and a 
particle reservoir as long as there is sufficient energy to keep a 
constant temperature. This means that under the constant pressure 
condition an explicit thermostat keeps a constant temperature despite 
imparting out (in) some amount of heat. The heat transmission only 
changes the volume fractions of two phases: the phase with higher heat 
capacity (for definiteness, a liquid) condenses a certain amount of gas 
under external cooling or it partly evaporates some amount of gas under 
external heating. Evidently, for finite systems the amount of 
transmitted heat is also finite and it depends on the masses of both 
phases and their heat capacities.  
Similarly, one can add to or remove from the system some amount of each 
phase, but under the constant pressure condition the system (mixed 
phase) will maintain a constant temperature and a full chemical 
equilibrium, i.e. the equal values of chemical potentials for both pure 
phases being in contact. Therefore, up to some maximal value any amount 
of the removed phase, namely the gas of hadrons, for definiteness, will 
be in a full chemical and thermal equilibrium with the mixed phase. 
Apparently, the maximal amount of removed phase depends on the masses of 
both phases and their energy densities. However, to justify such 
arguments it is necessary to assume that the mixed phase has 
sufficiently large volume, so the finite size effects are not important. 

The generalized shock adiabat model of central HIC 
\cite{KAB:89a,KAB:90,Bugaev:2015,Bugaev:2016a,Bugaev:2016b} allows one 
to reliably determine the initial conditions for the subsequent 
hydrodynamic evolution. As it was shown numerically in 
Refs.~\cite{Bugaev:2015,Bugaev:2016a,Bugaev:2016b} the initial states 
which correspond to a mixed phase have practically the same pressure 
and, hence, the condition of constant pressure is valid with the 
accuracy of about two percent \cite{Bugaev:2017a}. Therefore, the 
strangeness equilibrium dale seen at the collision energy 
$\sqrt s_{NN} = 4.9$~GeV is a new and  an independent signal of the 
mixed phase formation in HIC. Since the number of the effective degrees 
of freedom of almost massless particles estimated in \cite{Bugaev:2017a} 
is about 1770, it evidences for the CSR PT in hadronic phase at the 
collision energies  $\sqrt s_{NN} = 4.3-4.9$~GeV.  Furthermore, since 
above we came to a conclusion that at $\sqrt s_{NN} = 8.8-9.2$~GeV there 
is a deconfinement PT, it means that for the first time we have an 
experimental evidence that QCD may have a tricritical endpoint instead 
of a critical one.

In our opinion it is remarkable that an existence of two phase 
transitions in QCD is also supported by the most successful transport 
approach developed in Giessen \cite {Cassing16a,Cassing16b}. It is a 
well-known Parton-Hadron-String Dynamics transport model. 
Moreover, it seems that this approach predicts the CSR PT of first 
order at the collision energy about $\sqrt s_{NN} \simeq 4$~GeV, while 
the deconfinement PT of second order is predicted at about 
$\sqrt s_{NN} \simeq 10$~GeV \cite{Cassing16a,Cassing16b}. In contrast 
to the PT signals presented above, the transport model of 
Refs.~\cite{Cassing16a,Cassing16b} demonstrates  two cross-overs due to 
the finite size of the system. 

\begin{figure}[t]
% Use the relevant command for your figure-insertion program
% to insert the figure file.
\centering
\mbox{\includegraphics[width=69mm,height=63mm,clip]{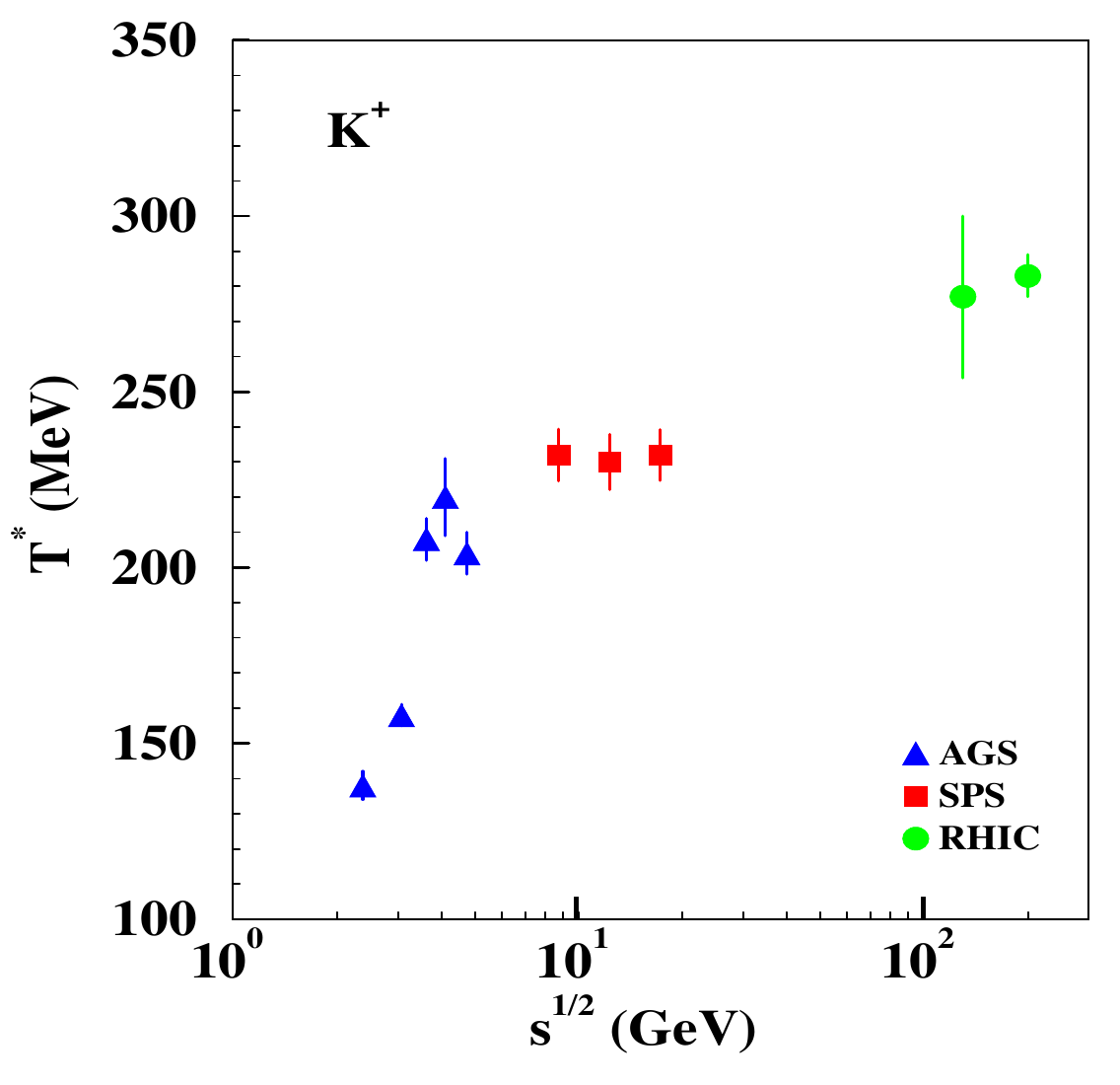}}
\caption{The collision energy dependence of the inverse slope parameter 
of $K^+$ mesons. The two topmost AGS points (triangles) demonstrate the 
irregular behavior which may be related to the CSR PT. This plot is 
taken from Ref.~\cite{Step}.}
\label{Bugaev_fig4}       % Give a unique label
\end{figure}

It is possible that the CSR is responsible for the decrease of the 
inverse slope parameter of $K^+$ mesons, when collision energy increases 
from $\sqrt s_{NN} = 4.3$~GeV to $\sqrt s_{NN} = 4.9$~GeV. As one can 
see from the two topmost triangles in Fig.~\ref{Bugaev_fig4} the inverse 
slope $T^*$, indeed, decreases. Of course, this maybe a coincidence, but 
on the other hand this can be also a manifestation of the CSR. For the 
transverse momentum spectra of particles of mass $m_k$ which have the 
mean transverse hydrodynamic velocity $\overline{v}_T$ and temperature 
$T$ one can get the formula \cite{SlopeT}
\begin{equation}\label{EqXII}
T^*_k (p_T\rightarrow 0)~=~\frac{T}{1~-~\frac{1}{2}~\overline{v}_T^2~
(m_k/T~-~1)}~\approx~ T~+ \frac{1}{2}~m_k~\overline{v}_T^2~,
\end{equation}
where $p_T$ is the transverse momentum of particle. Since it is hard to 
imagine that an increase of collision energy can lead to a decrease of 
the hadronization temperature $T$ or to a decrease of the mean 
transverse hydrodynamic velocity $\overline{v}_T$, then the only 
possible cause of the decrease of $T^*_k$ for $K^+$ mesons is that their 
mass is reduced. It is interesting that NA49 Collaboration also reported 
a similar change of the inverse slope parameter of  $K^-$ mesons,  but 
at a slightly higher collision energy interval $\sqrt s_{NN} = 
4.9-6.3$~GeV (see the left panel of Fig.~5 in \cite{StepPP}).
Therefore, in order to verify or to disprove our hypothesis it would be 
necessary to measure the inverse slope parameter of $K^\pm$ mesons (or 
their transverse masses) with high precision in the collision energy 
range $\sqrt s_{NN} = 4.3-6.3$~GeV.

\section{Conclusions}

From the discussions above one can unambiguously conclude that the IST 
EoS is perfectly suited to determine the hard-core radii of all hadrons 
from the hadronic multiplicities which will be measured in the future 
experiments on RHIC, NICA and FAIR. We hope that these experiments will 
help to verify the new signals of the CSR PT and the deconfinement PT 
outlined here, and to experimentally locate the tricritical endpoint of 
the QCD phase diagram.  

\begin{acknowledgement} 
The authors are thankful to D. B. Blaschke, R. A. Lacey,  
I. N. Mishustin and R. D. Pisarski for fruitful discussions and valuable comments. K.A.B., 
A.I.I., V.V.S. and G.M.Z. acknowledge the support of the National 
Academy of Sciences  of Ukraine. The work of K.A.B. and L.V.B. was 
performed in the framework of COST Action CA15213 ``Theory of hot matter 
and relativistic heavy-ion collisions" (THOR). K.A.B. is thankful to the  
COST Action CA15213 for a partial support. V.V.S. thanks the 
Funda\c c\~ao para a Ci\^encia e Tecnologia (FCT), Portugal, for the
financial support to make research at the Centro de Astrof\'{\i}sica e 
Gravita\c c\~ao (CENTRA), Instituto Superior T\'ecnico, Universidade de 
Lisboa. The work of L.V.B. and E.E.Z. was supported by the Norwegian 
Research Council (NFR) under grant No. 255253/F50 - CERN Heavy Ion 
Theory. K.A.B. is very thankful to the organizers of the ICNFP2017 for 
the warm hospitality in OAK. 
\end{acknowledgement}

%
% BibTeX or Biber users please use (the style is already called in the class, ensure that the "woc.bst" style is in your local directory)
% \bibliography{name or your bibliography database}
%
% Non-BibTeX users please use
%

\end{document}